\documentclass[aps,prl,twocolumn,letterpaper,nobibnotes,superscriptaddress,longbibliography]{revtex4-1}
\usepackage{amssymb}
\usepackage{amsmath}
\usepackage{graphicx}
\usepackage{extarrows}
\usepackage{url}
\usepackage{varioref}
\usepackage{hyperref}
\usepackage{cleveref}
\usepackage{color}
\usepackage{float}

\newcommand\ket[1]{\ensuremath{|#1\rangle}}

\begin{document}
\title{Creation of 2000-atom Greenberger-Horne-Zeilinger states by entanglement amplification}
\author{Yajuan Zhao}
\affiliation{Department of Physics and State Key Laboratory of Low Dimensional Quantum Physics, Tsinghua University, Beijing, 100084, China}

\author{Rui Zhang}
\affiliation{Department of Physics and State Key Laboratory of Low Dimensional Quantum Physics, Tsinghua University, Beijing, 100084, China}

\author{Wenlan Chen}
\email{To whom correspondence should be addressed to: ultracold@ultracold.cn (W. Chen and J. Hu) or xbwang@mail.tsinghua.edu.cn (X.-B. Wang).}
\affiliation{Department of Physics and State Key Laboratory of Low Dimensional Quantum Physics, Tsinghua University, Beijing, 100084, China}
\affiliation{Frontier Science Center for Quantum Information, Beijing, 100084, China}

\author{Xiang-Bin Wang}
\email{To whom correspondence should be addressed to: ultracold@ultracold.cn (W. Chen and J. Hu) or xbwang@mail.tsinghua.edu.cn (X.-B. Wang).}
\affiliation{Department of Physics and State Key Laboratory of Low Dimensional Quantum Physics, Tsinghua University, Beijing, 100084, China}
\affiliation{Frontier Science Center for Quantum Information, Beijing, 100084, China}
\affiliation{Synergetic Innovation Center of Quantum Information and Quantum Physics,
University of Science and Technology of China, Hefei, Anhui, 230026, China}
\affiliation{Jinan Institute of Quantum technology, SAICT, Jinan, 250101, China}
\affiliation{Shenzhen Institute for Quantum Science and Engineering, and Physics Department, Southern University of Science and Technology, Shenzhen, 518055, China}

\author{Jiazhong Hu}
\email{To whom correspondence should be addressed to: ultracold@ultracold.cn (W. Chen and J. Hu) or xbwang@mail.tsinghua.edu.cn (X.-B. Wang).}
\affiliation{Department of Physics and State Key Laboratory of Low Dimensional Quantum Physics, Tsinghua University, Beijing, 100084, China}
\affiliation{Frontier Science Center for Quantum Information, Beijing, 100084, China}

\begin{abstract}
We propose a novel entanglement-creation scheme in a multi-atom ensemble, named \textit{entanglement amplification}, which converts unentangled states into entangled states and amplifies less-entangled ones to maximally-entangled Greenberger-Horne-Zeilinger (GHZ) states. The scheme starts with a multi-atom ensemble initialized in a coherent spin state.
By shifting the energy of a particular Dicke state, we break the Hilbert space of the ensemble into two isolated subspaces to tear the coherent spin state into two components so that entanglement is introduced.
After that, we utilize the isolated subspaces to further enhance the entanglement by coherently separating the two components. By single-particle Rabi drivings on atoms in a high-finesse optical cavity illuminated by a single-frequency light, 2000-atom GHZ states can be created with a fidelity above 80\% in an experimentally achievable system, making resources of ensembles at Heisenberg limit practically available for quantum metrology.
\end{abstract}

\maketitle

Entanglement plays a central role in quantum mechanics. It is one of the most important topics in fields including quantum information \cite{WANG20071,RevModPhys.87.307,RevModPhys.80.1083}, quantum communication \cite{Kimble2008,Duan2001} and quantum metrology \cite{RevModPhys.90.035005,PhysRevA.47.5138,Zou6381}. By utilizing different classes of entangled states, one can speed up computations \cite{PhysRevLett.85.1334,PhysRevA.57.120,PhysRevLett.92.127902}, secure private communications \cite{Kuzmich2003,Ren2017,Sun2016,PhysRevLett.120.030501,PhysRevLett.119.200501}, and overcome the standard quantum limit \cite{PhysRevLett.104.013602,Appel10960,Gross2010,Riedel2010,PhysRevLett.104.073604,Bohnet2014,PhysRevLett.104.250801,PhysRevLett.111.103601} to get higher precision. Among all the classes of entangled states, the Greenberger-Horne-Zeilinger (GHZ) state \cite{GHZ} is one of the ultimate goals for quantum information and quantum metrology \cite{Leibfried1476,PhysRevLett.106.130506,Roos1478,Sackett2000,Lu2007,PhysRevLett.120.260502,PhysRevLett.121.250505,PhysRevLett.122.110501,PhysRevLett.119.180511,wei2019verifying,Song574,Omran2019}, for it displays the Heisenberg limit \cite{PhysRevA.54.R4649} with the best precision guaranteed by fundamental principles of quantum mechanics.

However, it is non-trivial or even challenging to create GHZ states in multi-particle ensembles. In the past few years, pioneering contributions have been made to realize multi-particle GHZ states at different platforms, including 14 trapped ions \cite{Leibfried1476,Roos1478,Sackett2000,PhysRevLett.106.130506}, 18 state-of-the-art photon qubits \cite{Lu2007,PhysRevLett.120.260502,PhysRevLett.121.250505}, and 12 superconducting qubits \cite{PhysRevLett.119.180511,PhysRevLett.122.110501}. These outstanding works start a new era in developing scalable quantum computers, advancing quantum metrology, and establishing quantum communication and teleportation. Recently there is a breakthrough where up to 20 qubits \cite{Omran2019,wei2019verifying,Song574} are entangled with a fidelity above 0.5 \cite{Sackett2000}. 
Nevertheless, the required precision of the control and technical difficulties increase exponentially as the number of qubits grows, making it difficult to increase the size of GHZ states.

\begin{figure}[htbp]
\begin{center}
\includegraphics[width=3.3in]{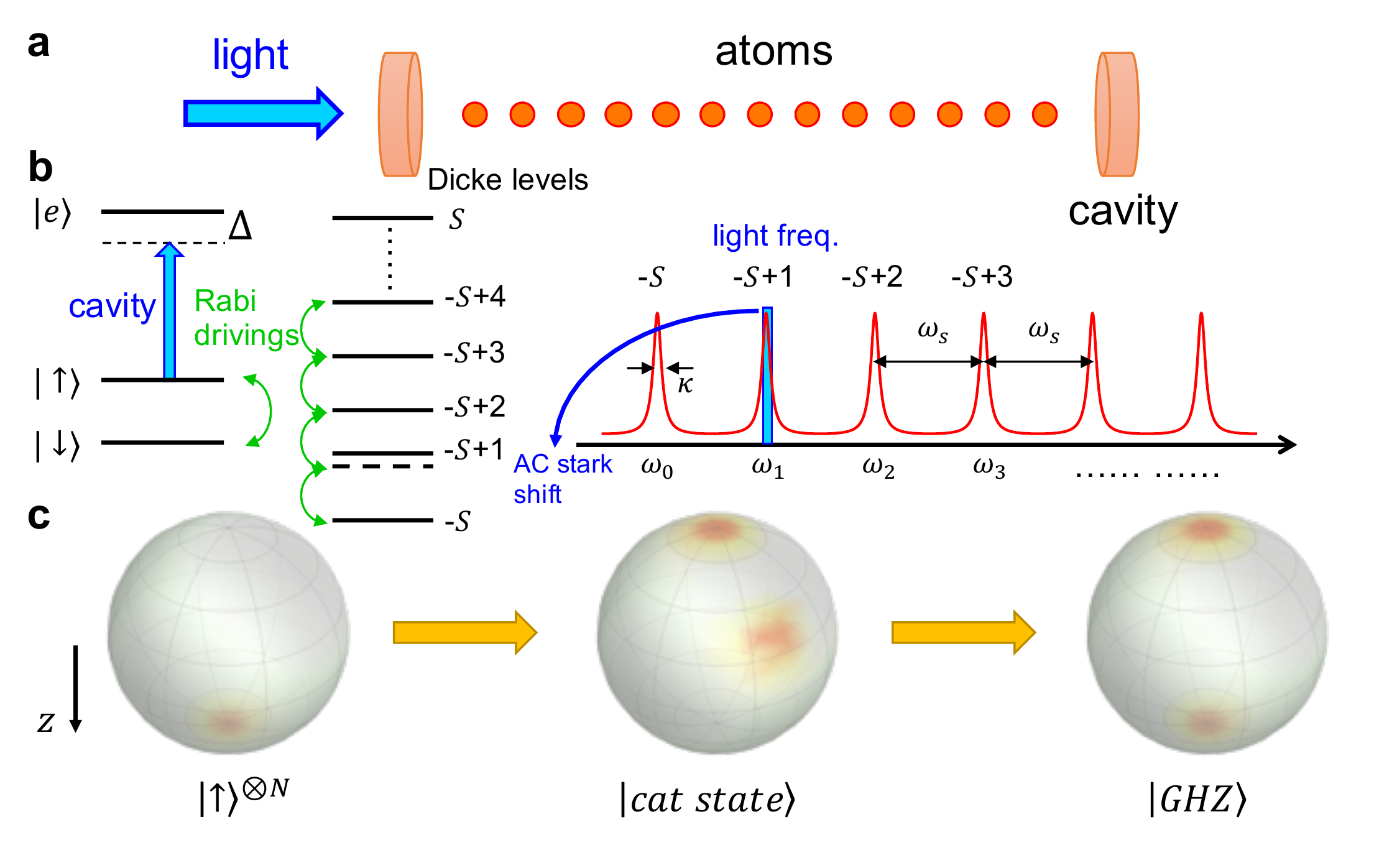}
\caption{Setup for \textit{entanglement amplification} in atomic ensembles. (\textbf{a}) $N$ atoms are coupled to a high finesse optical cavity, and the cavity is illuminated by a single-frequency light which can be turned on or off. (\textbf{b}) The atomic level diagram and the cavity transmission spectra. The Rabi drivings are applied to couple $\ket{\uparrow}$ and $\ket{\downarrow}$. The cavity mode couples $\ket{\uparrow}$ to $\ket{e}$ with a detuning $\Delta$. Due to the strong coupling in the atom-cavity system, each atom in $\ket{\uparrow}$ shifts the cavity resonance by an amount of $\omega_s=g^2/\Delta>\kappa$. Thus, the incident light with frequency of $\omega_1=\omega_c+\omega_s$ only shifts the energy of the Dicke state $\ket{-S+1}$ and creates a boundary in the Hilbert space at this Dicke state. (\textbf{c}) The quasi-probability distribution (Husimi-Q function) on the Bloch sphere before, within, and after \textit{entanglement amplification} process.}
\label{figure1}
\end{center}
\end{figure}

In this Letter, we propose a deterministic scheme, named \textit{entanglement amplification}, to convert non-entangled states into less-entangled states, and further amplify the less-entangled ones to maximally-entangled GHZ states in atomic ensembles. By shifting the energy of one particular angular momentum eigenstate of collective atomic spins (Dicke state \cite{PhysRev.93.99}), the Hilbert space is broken into two isolated subspaces separated by this energy-shifted boundary. Any wavefunction in one of the subspaces is not allowed to leak out to or penetrate from the other. When a coherent spin state is approaching the boundary by Rabi drivings between two spins of each atom, the wavefunction evolves around the boundary, being torn into two separated components, and finally becomes a cat state. 
Furthermore, by carefully choosing the subspace boundary and the orientation of the wavefunction, one component can be frozen, 
while the other continues rotating under Rabi drivings, which further stretches the wavefunction separation of the cat state, until the maximally-separated state (GHZ state) is obtained.
Estimated with experimentally achievable parameters, 
a 100-atom GHZ state can be obtained with a fidelity at 0.92, and the one with 2000 atoms can be achieved with a fidelity at 0.89. Moreover, we find the fidelity of GHZ states obtained using \textit{entanglement amplification} decreases logarithmically as the atom number increases, making it possible to extend this scheme into the regime of larger atom number.

We consider $N$ three-level atoms trapped in an optical cavity (see Fig.~\ref{figure1}), with two ground states $\ket{\uparrow}$ and $\ket{\downarrow}$, and one excited state $\ket{e}$. The cavity mode couples the state $\ket{\uparrow}$ to the state $\ket{e}$ with a single-photon Rabi frequency $2g$ and a detuning $\Delta$, where $\Delta$ is much larger than the spontaneous decay rate $\Gamma$ of the state $\ket{e}$. By adiabatically eliminating the state $\ket{e}$, we obtain an effective Hamiltonian $H_c$ \cite{PhysRevLett.115.250502,TANJISUZUKI2011201}, describing the interaction between the cavity field and $N$ two-level atoms:
\begin{equation}
H_c=\hbar \omega_s(S_z+S)\hat c^\dagger \hat c.
\end{equation}
Here, $S_z$ is the collective angular momentum operator along $z$ axis, $ \omega_s=g^2/\Delta$ is the coupling strength, $S=N/2$ is the total spin magnitude, and $\hat c^\dagger$ ($\hat c$) is the creation (annihilation) operator of the cavity field.

Each atom in the state $\ket{\uparrow}$ shifts the cavity resonance $\omega_c$ by an amount of $\omega_s$.
When the cavity is illuminated by a light beam at frequency $\omega_n=\omega_c+n\omega_s$, the intra-cavity intensity $\langle \hat c^\dagger \hat c\rangle_{m,n}$ is negligibly small if $m\neq n$, where $m$ is the number of atoms in the state $\ket{\uparrow}$. In this case, only quantum states with $n$ atoms in $\ket{\uparrow}$ introduce significant intra-cavity intensity, and thus introduce a significant AC stark shift $n\hbar\omega_s  \langle \hat c^\dagger \hat c\rangle_{n,n}$ to the Dicke state $\ket{m=-N/2+n}$, while the light-induced energy shifts of other Dicke states are negligible.
This achieves the goal of shifting one particular Dicke state away without affecting the others and thus forms a boundary separating the Hilbert space. In the following context, we choose an incident light beam at frequency $\omega_1=\omega_c+\omega_s$ to illuminate the cavity so that the boundary separating the Hilbert space is set to the Dicke state $\ket{m=-N/2+1}$. As a result, in an ideal case, the effective Hamiltonian becomes $H'(\delta)=\text{diag}(0,0,\ldots,\hbar\delta,0)$, where each diagonal matrix element $H'_{m,m}$ corresponds to the energy shift of the Dicke state $\ket{m=-N/2+n}$, as $n=N$ to $0$ in a descending order, and $\delta=\omega_s\langle \hat{c}^\dagger\hat{c}\rangle_{1,1}$.

\begin{figure}[t]
\begin{center}
\includegraphics[width=3.8in]{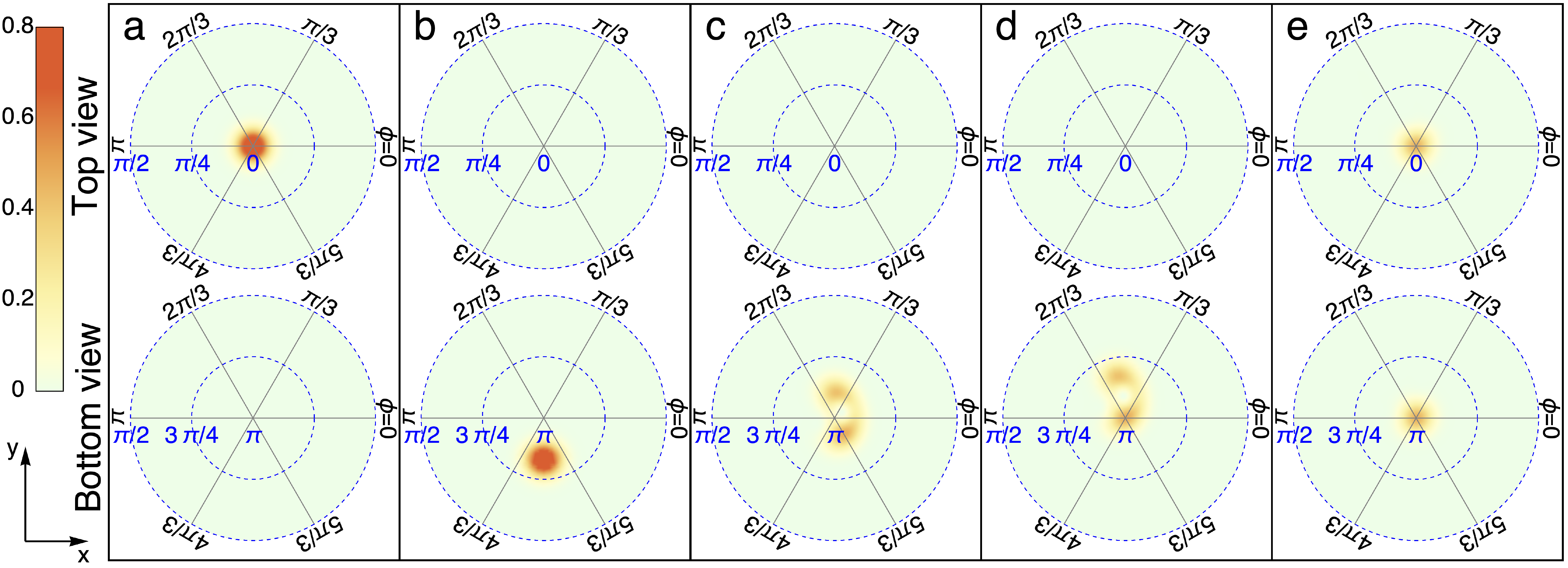}
\caption{
Four steps of \textit{entanglement amplification} to create a GHZ state from a coherent spin state. Here we plot the Husimi-Q function for a 100-atom ensemble by using the spherical coordinates $(\theta,\phi)$ on the Bloch sphere, where $\theta$ varies from 0 to $\pi$ and $\phi$ varies from 0 to $2\pi$. Each panel contains one top view ($\theta$ from 0 to $\pi/2$) and one bottom view ($\theta$ from $\pi/2$ to $\pi$) of the Bloch sphere. (\textbf{a}) A coherent spin state is initialized in $\ket{\uparrow}^{\otimes N}$. (\textbf{b}) After Step 1, the state is rotated close to $\ket{-N/2+1}$. (\textbf{c}) After Step 2, a cat state is created by the boundary at $\ket{-N/2+1}$. (\textbf{d}) After Step 3, the orientation of the cat state is re-aligned so that one component sits at the south pole of the Bloch sphere. (\textbf{e}) A GHZ state is created after Step 4 by freezing one component of the wavefunction in $\ket{-N/2}$ using the boundary of $\ket{-N/2+1}$.}
\label{figure2}
\end{center}
\end{figure}

We realize \textit{entanglement amplification} in the following steps.
Step 1: All the atoms are initialized in $\ket{\uparrow}$ and then rotated along $x$ axis by Rabi drivings approaching the Dicke state $\ket{m=-N/2+1}$ without turning on the incident light onto the cavity (Fig.~\ref{figure2}\textbf{a}$\rightarrow$\ref{figure2}\textbf{b}). This process can be described by the rotation Hamiltonian $\hbar\Omega S_x$ where $S_x$ is the collective angular momentum operator along $x$ axis and $\Omega$ is the Rabi frequency of single-particle Rabi drivings. Step 2: Turn on the cavity light to introduce the energy shift at Dicke state $\ket{m=-N/2+1}$, and continue the state rotation along $x$ axis (Fig.~\ref{figure2}\textbf{b}$\rightarrow$\ref{figure2}\textbf{c}). This process is described by $\hbar \Omega S_x+ H'$. Here we require $\sqrt{N}\Omega<|\delta|$ to guarantee the off-resonance condition. The wavefunction propagates around the Dicke state $\ket{m=-N/2+1}$ and evolves into two separate components. By choosing a proper time to stop applying such Rabi drivings, the ensemble evolves into a cat state $|\psi_{cat}\rangle$ where two components of the wavefunction are coherently separated on the Bloch sphere (Fig.~\ref{figure2}\textbf{c}):
\begin{equation}\label{catstate}
|\psi_{cat}\rangle=e^{-i\left[\Omega S_x+H'(\delta_2)/\hbar\right]t_2}e^{-i\Omega S_x t_1}|\uparrow\rangle^{\otimes N}, 
\end{equation}
where $t_1$ and $t_2$ represents the duration time of Step 1 and Step 2 respectively.

Then, we convert the obtained cat state into a GHZ state by two additional steps. Step 3: Turn off the cavity light, apply Rabi drivings to rotate the cat state until one component of the state is aligned into the south pole of the Bloch sphere described by the Dicke state $\ket{m=-N/2}$ (Fig.~\ref{figure2}\textbf{c}$\rightarrow$\ref{figure2}\textbf{d}). In this step, the separation between two components stays unchanged. Step 4: Turn on both the cavity light and the Rabi Drivings (Fig.~\ref{figure2}\textbf{d}$\rightarrow$\ref{figure2}\textbf{e}). The component in $\ket{m=-N/2}$ is frozen by the boundary at Dicke state $\ket{m=-N/2+1}$, while the other component is rotated into the state $\ket{m=N/2}$. A GHZ state $|\psi_{GHZ}\rangle$ with two coherent components each on the north and south pole of the Bloch sphere is thus obtained:
\begin{eqnarray}
|\psi_{GHZ}\rangle&=&e^{-i\left[\Omega S_{\phi_4} +H'(\delta_4)/\hbar\right]t_4}e^{-i\Omega S_{\phi_3} t_3}|\psi_{cat}\rangle, \label{evolution}
\end{eqnarray}
where $S_{\phi_i}=S_x\cos\phi_i+S_y\sin\phi_i$. Here, $e^{-i\Omega S_{\phi_3} t_3}$ corresponds to Step 3, the process of the orientation alignment, and $e^{ -i\left[\Omega S_{\phi_4} +H'(\delta_4)/\hbar\right]t_4 }$ corresponds to Step 4, the process of amplifying the entanglement.

With all these operations, we convert a non-entangled CSS into a less-entangled cat state, and amplify this cat state into a maximally-entangled GHZ state, assuming that the cavity lines are infinitely narrow compared with the amount of cavity resonance shift introduced by one atom in the ground state $|\uparrow\rangle$.  This situation corresponds to an infinitely large cavity cooperativity $\eta$.

In a realistic system with a finite cavity cooperativity, we need to consider dissipation due to spontaneous decay and effects of the finite cavity linewidth which lead to redistribution of the wavefunction among  two isolated Hilbert subspaces. Such processes will decrease the fidelity of the obtained GHZ state. 
Suppose the excited state $\ket{e}$ has a spontaneous decay rate $\Gamma$, and a detuned coupling from $\ket{\uparrow}$ to $\ket{e}$ brings an AC Stark shift $ E_{\uparrow}$ to the energy of $\ket{\uparrow}$, and then introduces a spontaneous decay rate $\Gamma_{\uparrow}= E_{\uparrow}\times\Gamma/\Delta$ for each atom in $\ket{\uparrow}$. 
After the 4-step operations, the density matrix of the atomic ensemble can be decomposed into two parts, the coherently-evolved part and the incoherently-scattered part. Since the latter part can be described by a positive-defined density matrix, it should contribute a non-negative number to the fidelity of the obtained GHZ state. Thus, the fidelity estimated only by the coherent-evolved density matrix provides a lower bound for the fidelity of the obtained GHZ state. Without losing generality, we choose to name this lower bound as fidelity $\mathcal F$ in the following context for simplicity.

When the cavity has a finite linewidth, the cavity light at frequency $\omega_1$ introduces non-negligible AC Stark shifts to other Dicke states besides $\ket{m=-N/2+1}$.
Therefore, in a realistic system, a modified non-Hermitian Hamiltonian $ H'_{\text{exp}}$ replaces the ideal $ H'$ to describe the cavity linewidth broadening and the cavity-assisted energy shift under the dissipation of spontaneous decay (see Supplemental Information \cite{SI}):
\begin{eqnarray}
 H'_{\text{exp}}&=&{\hbar\delta\left(1-i{\Gamma\over2\Delta}\right)\over|T(\omega_s,1)|^2}\times\text{diag}  (N|T(\omega_s,N)|^2,\nonumber \\
& &(N-1)|T(\omega_s,N-1)|^2, \ldots,\nonumber \\
& &1\times|T(\omega_s,1)|^2,0\times|T(\omega_s,0)|^2 ) .\label{Hexp}
\end{eqnarray}
The real part of $ H'_{\text{exp}}$ characterizes AC Stark shifts for different Dicke states and the imaginary part characterizes the spontaneous-decay-induced decoherence. Here, $T(\xi,n)$ is the amplitude transmission function of the cavity \cite{TANJISUZUKI2011201}:
\begin{equation}
T(\xi,n)={1\over 1+ {n\eta\over 1+4(\Delta+\xi)^2/\Gamma^2}-2i\left[{\xi\over\kappa}-n\eta {(\Delta+\xi)/\Gamma \over 1+4(\Delta+\xi )^2/\Gamma^2}\right]}, \label{transmission}
\end{equation}
where $n$ is the atom number in the state $\ket{\uparrow}$, $\eta=4g^2/\left(\Gamma\kappa\right)$ is the cavity cooperativity, $\kappa$ is the linewidth of the cavity, and $\xi=\omega-\omega_c$ is the light-cavity detuning.

To verify the validity of our scheme, we use experimentally achievable parameters to estimate the fidelity of the achieved GHZ state. We consider rubidium-87 as the candidate atom, with two ground states $|\downarrow\rangle$ and $|\uparrow\rangle$ in different hyperfine manifolds of $5S_{1/2}$, and a excited state $|e \rangle$ in $5P_{3/2}$ with a spontaneous decay rate $\Gamma=2\pi\times 6~$MHz. Choosing cavity cooperativity $\eta=200$, and Rabi frequency of single-particle Rabi drivings $\Omega$ between $2\pi\times 0.05$~MHz and $2\pi\times 0.2$~MHz (see SI \cite{SI} for details of all parameters), a GHZ state with a fidelity of 0.92 is achieved in a 100-atom ensemble, and a fidelity of 0.89 is achieved in a 2000-atom ensemble.  

\begin{figure}[t]
\begin{center}
\includegraphics[width=3in]{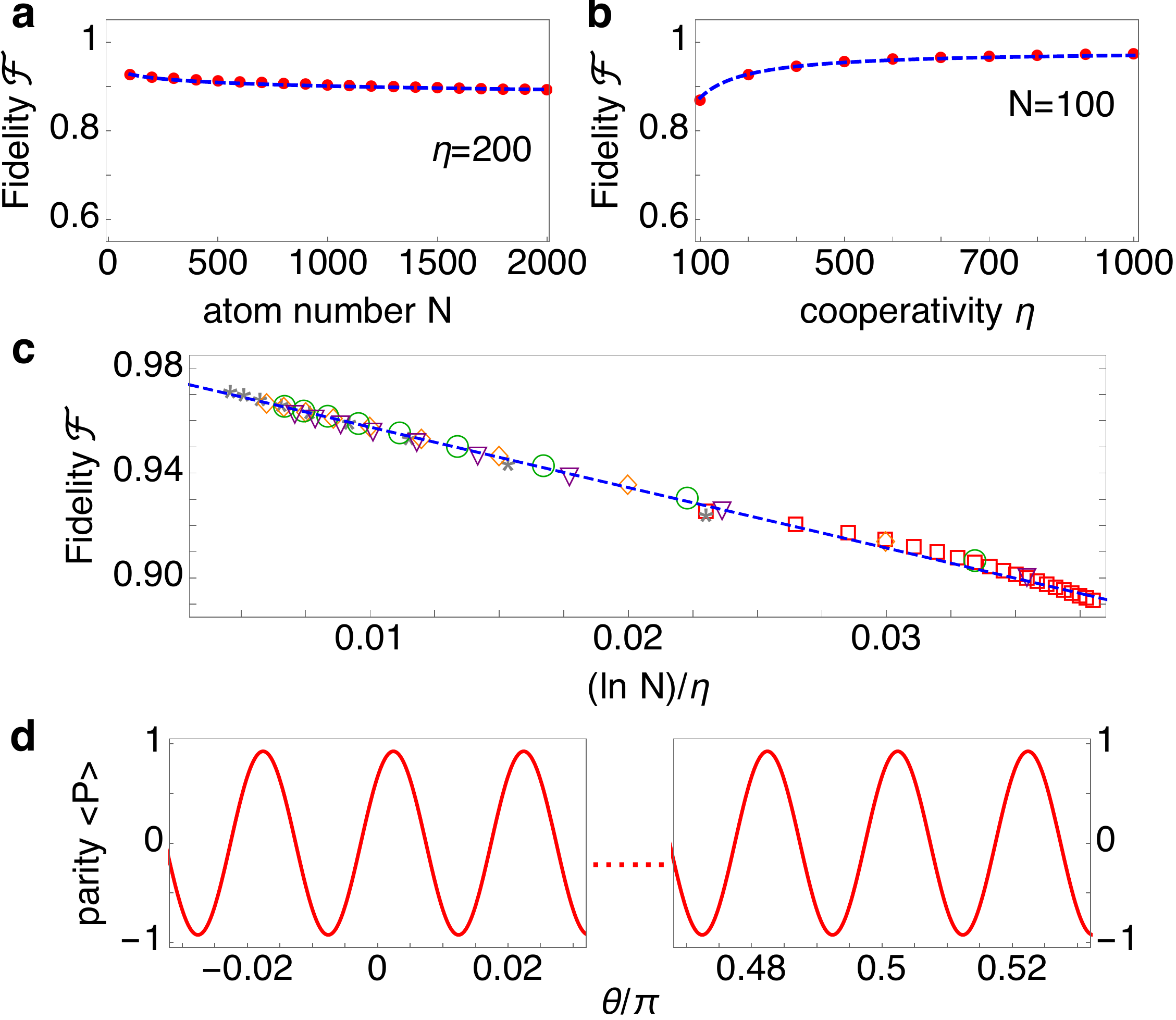}
\caption{Characterization of obtained GHZ states. (\textbf{a}) The fidelity $\mathcal F$ versus the atom number $N$ at $\eta=200$. 
(\textbf{b}) $\mathcal F$ versus the cooperativity $\eta$ at $N=100$. The dashed blue lines shows the empirical formula of $\mathcal F$ in Eq. \ref{Ffor}.
(\textbf{c}) $\mathcal F$ versus $(\ln N)/\eta$. Here we plot five sets of calculation results. 
The red squares correspond to a fixed $\eta=200$ with different atom number from $N=100$ to 2000. 
The gray stars correspond to $N=100$ with different $\eta$ from 200 to 1000. 
The orange diamonds correspond to $N=400$ with $\eta=200$ to 1000. 
The green circles correspond to $N=800$ with $\eta=200$ to 1000. 
The purple triangles  correspond to $N=1200$ with $\eta=200$ to 1000. 
The blue dashed line corresponds to the empirical formula of $\mathcal F$.
(\textbf{d}) The parity oscillation of $\langle P\rangle$ versus the rotating angle $\theta$. Here we only plot two typical intervals $[-0.03\pi,0.03\pi]$ and $[0.47\pi,0.53\pi]$ while the rapid oscillation of $\langle P\rangle$ is within the whole region of $\theta\in[-\pi,\pi]$.}
\label{figure3}
\end{center}
\end{figure}

The fidelity $\mathcal F$ of the obtained GHZ state has favorable scaling on atom number $N$, as plotted in Fig.~\ref{figure3}\textbf{a}. When atom number $N$ increases, the fidelity $\mathcal F$ of the obtained GHZ state decreases due to the dissipation induced by spontaneous decay. For the atomic population in boundary state $\ket{m=-N/2+1}$ which has the highest spontaneous decay rate, the dissipation is suppressed due to little population at this boundary state resulted from off-resonant Rabi coupling.
For atomic population in the other states, the spontaneous decay rate is low because the incident light is off-resonantly suppressed by the cavity linewidth. Such dissipation is proportional to $1/n$ for atomic population in state $|m=-N/2+n\rangle$ for $n=2$ to $N$, and thus introduces an overall dissipation proportional to $\ln N$. This weak dependence of $\mathcal F$ on $N$ helps to extend the scheme into the regime of thousands of atoms.

To understand the dependence of fidelity $\mathcal F$ on cavity cooperativity $\eta$, we plot fidelities of obtained GHZ states with 100 atoms at different cavity cooperativity $\eta$ with corresponding optimized detuning $\Delta$ (Fig.~\ref{figure3}\textbf{b}). The optimized $\Delta$ is proportional to $\eta$, and the optimized $t_i$ are the same for different $\eta$. According to Eq. \ref{Hexp} and \ref{transmission}, we find when $\eta$ increases, the coherent evolution keeps unchanged because $\eta/\Delta$ is a constant, but the spontaneous decay rate decreases inversely proportional to $\eta$ (or $\Delta$). The empirical formula for the fidelity $\mathcal F$ versus $N$ and $\eta$ is (see Fig.~\ref{figure3}\textbf{c} and SI \cite{SI} for more discussions):
\begin{equation}
\mathcal F =0.98-2.3(\ln N)/\eta.\label{Ffor}
\end{equation}

The obtained GHZ states can be verified experimentally by detecting the parity oscillation \cite{Sackett2000}. In our case, we apply a rotation $e^{i\pi S_\theta/2}$ to the obtained GHZ state and then measure the mean value of the parity operator $P=\prod_{i=1}^N \sigma_z^{(i)}$, where $\sigma_z^{(i)}$ is the Pauli $z$-matrix of the $i$-th atom. The parity $\langle P\rangle \sim\cos\left(N\theta\right)$ oscillates versus $\theta$ (see Fig.~\ref{figure3}\textbf{d}), proving the non-trivial coherence of $N$ atoms between the states $\ket{m=-N/2}$ and $\ket{m=N/2}$, and thus the measured state is a GHZ state. 

\begin{figure}[t]
\begin{center}
\includegraphics[width=2.8in]{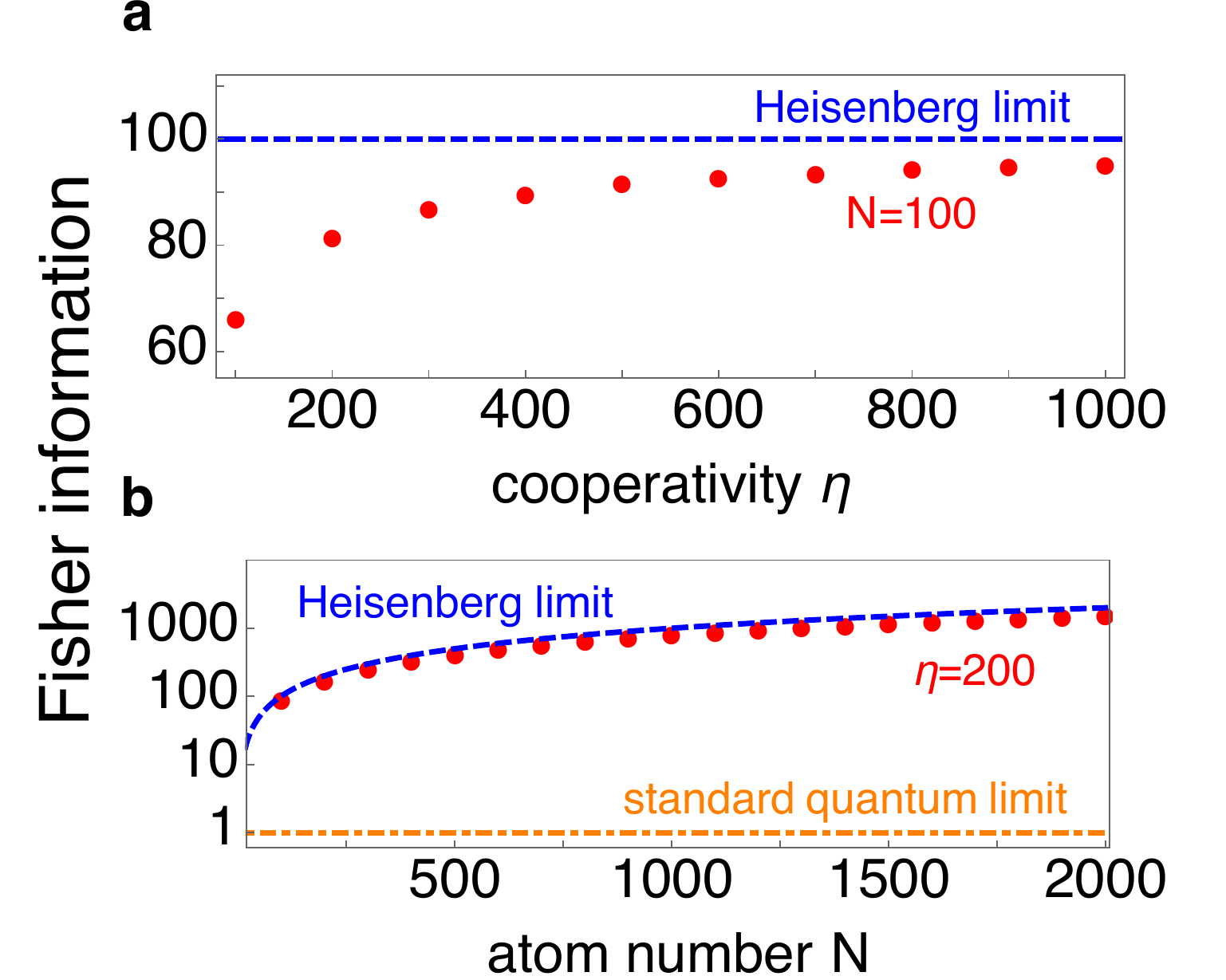}
\caption{Fisher information and metrological gain of the obtained GHZ states. Here we plot the normalized Fisher information relative to the CSS to characterize the metrological gain. (\textbf{a}) Fisher information of the generated GHZ states versus cavity cooperativity $\eta$ at  $N=100$ (red solid circles). (\textbf{b}) Fisher information of the generated GHZ states versus the atom number $N$ at $\eta=200$ (red solid circles). For reference, we also show the fisher information of states at the Heisenberg limit (blue dashed line) and those at the standard quantum limit (orange dot-dashed line).}
\label{figure4}
\end{center}
\end{figure}

To show that the obtained GHZ state is useful for metrological purposes, we plot its Fisher information which characterizes its metrological gain relative to that of a CSS (Fig.~\ref{figure4}). Here, we plot the Fisher information of the obtained GHZ states at different cooperativity $\eta$ (Fig.~\ref{figure4}\textbf{a}) and different atom number $N$ (Fig.~\ref{figure4}\textbf{b}). At a given $\eta=200$, the relative Fisher information reaches 81 for 100 atoms, 380 for 500 atoms, and 1420 for 2000 atoms while the relative Fisher information of a CSS is 1.
It confirms that \textit{entanglement amplification} strongly amplifies the metrological gain in a many-body system, approaching the Heisenberg limit at a given atom number $N$. 

Our method is also robust against common experimental noises (see SI \cite{SI} for detailed discussions). Considering the decoherence of the atomic states induced by magnetic field fluctuations, since the coherence time of an atomic clock with thousands of atoms could be on the order of seconds, a 2000-atom GHZ state should survive at least for milliseconds, which is long enough to finish all our processes  to generate the GHZ state that requires a timescale on the order of $2\pi/\Omega\approx10~\mu$s.
We also estimate the effects of other technical noises on the fidelity of the obtained GHZ state in SI \cite{SI}, including the precision of Rabi rotations, the atom-cavity inhomogeneous coupling, the photon shot noise, and the cavity frequency instability. The noises decrease the fidelity of the obtained GHZ state in two ways.
One way is to introduce errors in the rotation angles $\Omega t_i$ used in Eq. $\ref{catstate}$, $\ref{evolution}$. For a system containing $N$ atoms, the error of each rotation angle is required to be much smaller than the angle corresponds to the standard quantum limit $1/\sqrt N$, which has already been technically achieved in most of atomic clock apparatus. 
The other way is to introduce fluctuations on AC Stark shifts. The drifts or fluctuations of AC Stark shifts only bring significant influence on the resonant state $\ket{m=-N/2+1}$. Since such shift is only used to decouple the Rabi driving between $\ket{m=-N/2+1}$ and other states, the fidelity of the obtained GHZ state is not sensitive to the amount of the shift as long as the shift is large enough.
Taking all these potentially adverse conditions into consideration, we find the overall fidelity decreases to 0.920 (or 0.831) for $N=100$ (or $N=2000$), while the original value is 0.924 (or 0.890). This result further confirms the robustness of our scheme which could create GHZ states with an atom number as large as a few thousands.

The transmitted photons through the optical cavity serve as a measurement distinguishing whether the state is in $\ket{m=-N/2+1}$ or not. The photon number measurement may leak the information of the atomic states and introduce decoherence. This can be fixed by using a single-side cavity or an asymmetric cavity where the transmission of two mirrors are quite different. Using a cavity with finesse of $10^5$ and mirror transmission ratio of 0.09, the fidelity of the obtained GHZ state decreases by 0.06 due to information leakage. The detail discussions can be found in SI \cite{SI}.

In conclusion, we propose a new scheme, \textit{entanglement amplification}, for creating entangled states with high metrological gain. With realistic experimental parameters, one can obtain a 2000-atom GHZ state with a fidelity of 89\% and approach the Heisenberg limit. 
The fidelity decreases only logarithmically when the system size increases, which paves a new way to generate GHZ states with large size.
We believe this scheme simplifies the complexity and enhances the robustness of the creation of many-body entanglement. It may raise a new platform for designing simpler and more robust entanglement-creation schemes for quantum information and quantum metrology. 
Variations of this method can be generalized to artificial-atom systems such as superconducting qubits, quantum dots, and mechanical oscillators coupled to a resonator.

Y. Zhao, W. Chen and J. Hu acknowledge financial support from the National Natural Science Foundation of China under grant no. 11974202 and 61975092. R. Zhang and X.-B. Wang acknowledge financial support from Ministration of Science and Technology of China through The National Key Research and Development Program of China under grant no. 2017YFA0303901 and the National Natural Science Foundation of China under grant no. 11474182, 11774198 and U1738142.

\bibliography{GHZstate}

\setcounter{equation}{0}
\setcounter{figure}{0}
\setcounter{table}{0}

\widetext
\newpage
\renewcommand{\theequation}{S\arabic{equation}}
\renewcommand{\thefigure}{S\arabic{figure}}
\renewcommand{\thetable}{S\arabic{table}}
\Large
\begin{center}
\textbf{Supplemental information for Creation of 2000-atom}

\textbf{Greenberger-Horne-Zeilinger states by entanglement amplification}
\end{center}

\normalsize
In this supplemental information, we summarize the parameters used in numerical calculations
in the main text. We also analyze effects of different kinds of experimental noises on the fidelity of the obtained GHZ states.
This supplemental information serves as a support of our major conclusion that \textit{entanglement amplification} can create a GHZ state with thousands of atoms in an experimentally achievable system.
\tableofcontents

\section{The derivation of Hamiltonian $H'_{\text{exp}}$}
When the atom-cavity detuning $\Delta$ is large enough, we adiabatically eliminate the excited state $\ket{e}$ and obtain a Hamiltonian of the atom-cavity system:
\begin{equation}
H=\hbar\omega_s (S_z+\frac{N}{2}) c^\dagger c+\hbar\omega_c c^\dagger c+\hbar\omega_0 S_z+i\hbar B e^{i\omega t} c-i\hbar B e^{-i\omega t}c^\dagger.
\end{equation}
Here, $\hbar \omega_0$ is the energy gap
between the ground states $\ket{\uparrow}$ and $\ket{\downarrow}$, $\omega$ is the frequency of the incident light, and $B$ is a C-number that describes the incident light pumping the cavity field. We convert this Hamiltonian into interaction picture and get
\begin{equation} \label{S2}
H=\hbar \omega_s (S_z+\frac{N}{2}) c^\dagger c+ i\hbar B e^{i \xi t} c-i\hbar B e^{-i \xi t}c^\dagger.
\end{equation}
Here $\xi$ is the light-cavity detuning. By applying the Heisenberg-Langevin equation, the evolutions of the annihilation operator $c$ and the spin raising operator $S^+$ are described by
\begin{eqnarray}
i{d c\over d t}&=&-\kappa c/2 -i\omega_s (S_z+\frac{N}{2}) c-B e^{-i\xi t}, \nonumber \\
i{d S^+\over dt}&=&-\omega _s S^+ c^\dagger c .
\end{eqnarray}
By solving the equation above, we get $c(t)=c(0)e^{-\kappa t/2-i\omega_s S_z t}-{2B\over \kappa}{1\over 1+i{2\over \kappa}(\omega_s S_z-\xi)}e^{-i\xi t}$. Then we put this expression of $c(t)$ into Eq.~\ref{S2}. By averaging the initial vacuum state and eliminating the cavity field, we obtain
\begin{equation}
H=\hbar\omega_s (S_z+\frac{N}{2})\left(\frac{2B}{\kappa}\right)^2\frac{1}{1-\frac{4}{\kappa^2}(\omega_s S_z-\xi)^2}.
\end{equation}
This describes a state-dependent AC stark shift in the atom-cavity system. Here the factor $1/\left[ 1-{4\over \kappa^2}(\omega_s S_z-\xi)^2\right]$ fits
a Lorentzian shape with a linewidth $\kappa$ and a $S_z$-dependent central frequency, which
happens to be the same as the transmission spectrum for a symmetric cavity. To include the saturation and spontaneous decay of the atoms in this factor, we replace it with $|T(\xi,S_z+\frac{N}{2})|^2$ \cite{TANJISUZUKI2011201}. Considering decoherence induced by atomic spontaneous decay, we get the experimental Hamiltonian $H'_{exp}$:
\begin{equation}
H'_\text{exp}=\hbar\omega_s\left(1-i\frac{\Gamma}{2\Delta}\right)\left(\frac{2B}{\kappa}\right)^2 \left|T(\xi,S_z+\frac{N}{2})\right|^2\left(S_z+\frac{N}{2}\right),
\end{equation}
which can be modified into
the form of Eq. 4 in the main text.

\section{The analytical expression of state evolution and fidelity $\mathcal F$ of obtained GHZ states}
We replace $H'$ in Eq. 3 in the main text by the experimental Hamiltonian $H'_{\text{exp}}$. The experimentally obtained GHZ state can be described by
\begin{equation}
|\psi_{\text{GHZ}}\rangle=e^{-i\left[\Omega S_{\phi_4} +H'_{\text{exp}}(\delta_4)/\hbar\right]t_4}e^{-i\Omega S_{\phi_3} t_3} e^{-i\left[\Omega S_x+H'_{\text{exp}}(\delta_2)/\hbar\right]t_2}e^{-i\Omega S_x t_1}|\uparrow\rangle^{\otimes N}, \label{S1}
\end{equation}
where
\begin{equation}
 H'_{\text{exp}}(\delta)={\hbar\delta\left(1-i{\Gamma\over2\Delta}\right)\over|T(\omega_s,1)|^2}\text{diag}  (N|T(\omega_s,N)|^2,(N-1)|T(\omega_s,N-1)|^2, \ldots,1\times|T(\omega_s,1)|^2,0\times|T(\omega_s,0)|^2 ) \label{Hexp}
\end{equation}
and
\begin{equation}
T(\xi,n)={1\over 1+ {n\eta\over 1+4(\Delta+\xi)^2/\Gamma^2}-2i\left[{\xi\over\kappa}-n\eta {(\Delta+\xi)/\Gamma \over 1+4(\Delta+\xi )^2/\Gamma^2}\right]}. \label{transmission}
\end{equation}

Eq. \ref{S1} is also equivalent to applying rotations only along $x$ and $z$ axis, \textit{i.e.}
\begin{equation}
e^{i S_z \phi_4}|\psi_{\text{GHZ}}\rangle=e^{-i\left[\Omega S_{x} +H'_{\text{exp}}(\delta_4)/\hbar\right]t_4}e^{i S_z (\phi_4-\phi_3)} e^{-i\Omega S_{x} t_3}e^{i S_z \phi_3} e^{-i\left[\Omega S_x+H'_{\text{exp}}(\delta_2)/\hbar\right]t_2}e^{-i\Omega S_x t_1}|\uparrow\rangle^{\otimes N} \label{S4}
\end{equation}
which simplifies the expression in our numerical calculations.

The overall evolution with atomic spontaneous decay can be described by the master equations with Lindblad forms. The quantum fluctuation of the damping is smeared out by the ensemble average which suggests the final state to be a mixed state rather than a pure state.
We can use a density matrix $\rho=\rho_{coh}+\rho_{decay}$ which contains two parts to describe the obtained GHZ state: $\rho_{coh}=|\psi_{\text{GHZ}}\rangle\langle\psi_{\text{GHZ}}|$ represents the coherent-evolution part, while $\rho_{decay}$ corresponds to the incoherent-scattered part where the spontaneous decay introduces atom loss.
Because of the fragility of the GHZ state, any atom experiencing spontaneous decay will completely destroy the whole state. Thus, we are only interested in the coherent-evolution part and we define the fidelity by its lower bound 
\begin{equation}
\mathcal F=\max_\phi \langle \text{GHZ},\phi|\rho_{coh}|\text{GHZ},\phi \rangle,
\end{equation}
where $|\text{GHZ},\phi\rangle=(|\uparrow\rangle^{\otimes N}+e^{i\phi}|\downarrow\rangle^{\otimes N})/\sqrt{2}$. Here we use the phase $\phi$ to match the known phase difference between $|\uparrow\rangle^{\otimes N}$ and $|\downarrow\rangle^{\otimes N}$ of the generated GHZ state $|\psi_{\text{GHZ}}\rangle$.
We can also use the matrix elements $\rho_{m,n}$ to describe the fidelity $\mathcal F$:
\begin{equation}
\mathcal F={1\over 2}(\rho_{N/2,N/2}+\rho_{-N/2,-N/2}+|\rho_{-N/2,N/2}|+|\rho_{N/2,-N/2}|),
\end{equation}
where $\rho_{m,n}$ corresponds to the coefficients of $|m\rangle\langle n|$ in $\rho_{coh}$. This method was first used to characterize the fidelity of the GHZ state in Ref. \cite{Sackett2000}.

\section{The creation of a GHZ state with $N=100$ and $\eta=200$}

\begin{figure}[htbp]
\begin{center}
\includegraphics[width=3in]{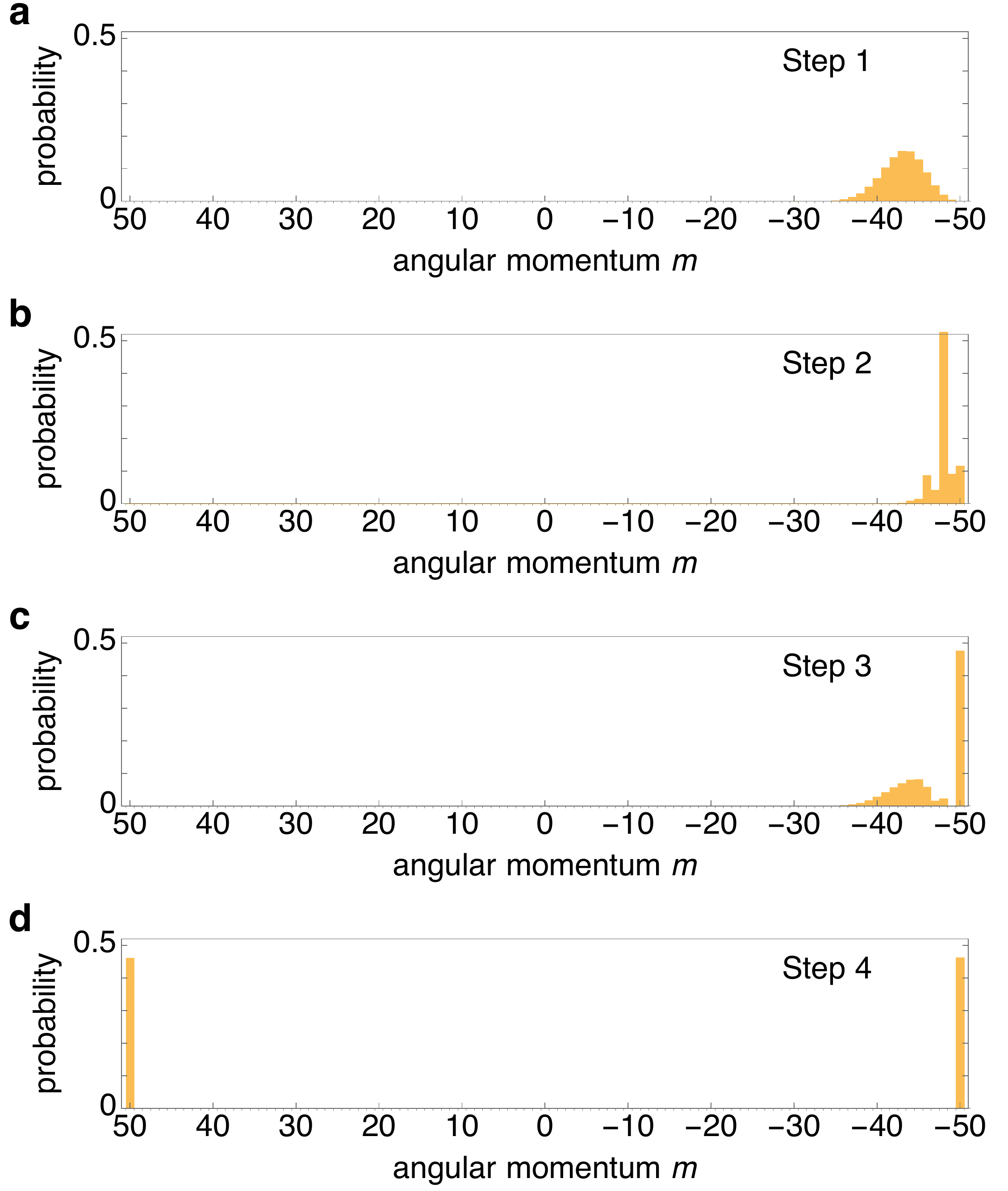}
\caption{The atomic population distribution on the basis of angular momentum eigenstates after Step 1 (\textbf{a}), 2 (\textbf{b}), 3 (\textbf{c}), or 4 (\textbf{d}).}
\label{sfig1}
\end{center}
\end{figure}

In this section, we concretely illustrate the creation and state evolution of a 100-atom GHZ state in a cavity with a cooperativity at $\eta=4g^2/(\kappa\Gamma)=200$.
The parameters we used to simulate an experimental case using Eq. \ref{S4} are listed below: $\kappa=2\pi\times 0.1$~MHz, $\Gamma=2\pi\times6$~MHz, $\Delta=-36\Gamma=-2\pi\times 216$~MHz, $\Omega=2\pi\times 0.2$~MHz, $t_1=2.074~\mu$s, $t_2=0.285~\mu$s, $t_3=0.191~\mu$s, $t_4=2.084~\mu$s, $\phi_{3}=0.503$, $\phi_{4}=0.257$, $\delta_2=-2\pi\times 4$~MHz, and $\delta_4=-2\pi\times 18.4$~MHz. We also plot the population distribution of atomic states on the angular momentum basis after each step in Fig. \ref{sfig1}.

In Step 1, 
the atomic ensemble is initialized in $|\uparrow\rangle^{\otimes N}$ and then rotated by an angle $-\Omega t_1$ along $x$ axis due to Rabi driving(Fig.~\ref{sfig1}\textbf{a}), \textit{i.e.}
\begin{equation}
|\text{Step 1}\rangle=e^{-i\Omega S_x t_1}|\uparrow\rangle^{\otimes N}.
\end{equation}

In Step 2, the incident light is turned on and we continue the Rabi driving along $x$ axis by an angle $-\Omega t_2$ (Fig.~\ref{sfig1}\textbf{b}). The evolution can be described by
\begin{equation}
|\text{Step 2}\rangle=e^{-i\left[\Omega S_x+H'_{\text{exp}}(\delta_2)/\hbar\right]t_2}|\text{Step 1}\rangle
\end{equation}

In Step 3, we realign the orientation of the wavefunction while turning off the incident light (Fig.~\ref{sfig1}\textbf{c}). The evolution is
\begin{equation}
|\text{Step 3}\rangle=e^{-i\Omega S_{\phi_3} t_3}|\text{Step 2}\rangle
\end{equation}

In Step 4, the incident light is turned on again but with larger light shift induced by higher light intensity, and we continue the rotation along the orientation of $\phi_{4}$ 
by an angle $-\Omega t_4$ (Fig.~\ref{sfig1}\textbf{d}). Therefore, the final obtained state in Step 4 is
\begin{equation}
|\psi_{\text{GHZ}}\rangle=|\text{Step 4}\rangle=
e^{-i\left[\Omega S_{\phi_4} +H'_{\text{exp}}(\delta_4)/\hbar\right]t_4}|\text{Step 3}\rangle
\end{equation}

The matrix elements can be obtained as $\rho_{N/2,N/2}=0.462$, $\rho_{-N/2,-N/2}=0.462$, and $\rho_{N/2,-N/2}=0.335+0.318i$. The fidelity $F$ of the obtained GHZ state equals 0.924.

\section{The numerical parameters used in Fig.~3\textbf{a} and \textbf{b}}

Here we summarize the specific numerical parameters used in Fig.~3\textbf{a} and \textbf{b}. In Fig.~3\textbf{a} in the main text, we calculate $\mathcal F$ versus the atom number $N$ at a given cooperativity $\eta=200$, and the parameters are summarized in Table~\ref{stable1}. In Fig.~3\textbf{b} in the main text, $\mathcal F$ is plotted versus $\eta$ at a given atom number $N=100$, and the parameters are summarized in Table~\ref{stable2}.

\begin{figure}[htbp]
\begin{center}
\includegraphics[width=7in]{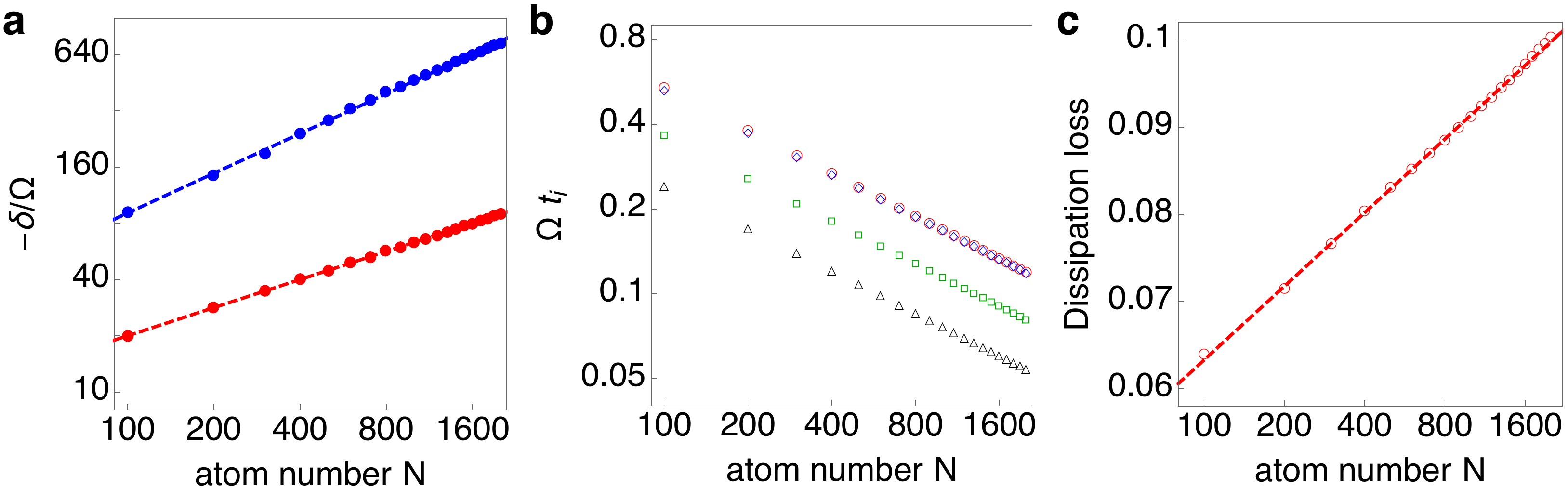}
\caption{\textbf{a}, The red and blue circles correspond to the calculation results of the relative light shift $|\delta_2/\Omega|$ and $|\delta_4/\Omega|$ respectively for different atom number $N$. The red and blue dashed lines are the linear fittings in log-log plot with slopes at 0.5 and 0.707. The calculation results indicate a power-law dependence between $|\delta_i/\Omega|$ and $N$ with a power order at $1/2$ and $1/\sqrt{2}$. \textbf{b}, The green squares, black triangles, red circles and blue diamonds correspond to the rotated angle $\Omega t_2$, $\Omega t_3$, $\pi-\Omega t_1$, and $\pi-\Omega t_4$ respectively. 
When fitted in the log-log plot, the slopes are -0.501 (green), -0.500 (black), -0.503 (red), and -0.497 (blue), which confirm a power-law dependence with a scaling at -1/2. \textbf{c}, we calculate the dissipation loss versus the atom number (red circles), and the red dashed line is its linear fit in the log-linear plot. It shows the probability of spontaneous decay is proportional to $\ln N$.}
\label{sfig2}
\end{center}
\end{figure}

Here we apply more quantitative analysis to Table~\ref{stable1}. We plot the relative light shift $|\delta_i/\Omega|$ ($i=2$ or $4$), rotated angles of each step, and the dissipation loss versus the atom number $N$ (Fig.~\ref{sfig2}). In Fig.~\ref{sfig2}\textbf{a}, we find the light shifts for Step 2($|\delta_2/\Omega|$) and Step 4 ($|\delta_4/\Omega|$) are proportional to $N^{1/2}$ and $N^{1/\sqrt{2}}$ respectively.
In Fig.~\ref{sfig2}\textbf{b}, the rotated angle $\Omega t_2$, $\Omega t_3$, $\pi-\Omega t_1$, and $\pi-\Omega t_4$ are proportional to $N^{-1/2}$. A possible explanation of this power-law dependence is that all physical process near the south pole of the Bloch sphere can be mapped into a flat plane by the Holstein-Primakoff transformation, where $S_y$ and $S_z$ can be considered as the conventional quadratures $x$ and $p$ in the quantum optics. The scaling of the mapping is proportional to $\sqrt{N}$. 
In the creation process of GHZ states, we need to deal with the curvature of the Bloch sphere. This is why the dependence of the light shift for Step 4
is optimized to be proportional to $N^{1/\sqrt{2}}$. 

In Fig.~\ref{sfig2}\textbf{c}, we calculate the probability of spontaneous decay $p_{decay}=1-\langle\psi_\text{GHZ}|\psi_\text{GHZ}\rangle$. We find $p_{decay}$ is proportional to $\ln N$ which indicates a very weak dependence on the atom number. Here we propose a qualitative argument to give a possible explanation.
By inspecting the imaginary parts of Eq.~\ref{Hexp}, we find the spontaneous decay rate for the state $\ket{-N/2+m}$ is approximately proportional to ${2\Delta\over m\eta^2 \Gamma}$. By summing up $m$ from 2 to $N$ for all Dicke states, we find
a scaling dependence $\ln N$, which means when the atom number $N$ increases, the probability of spontaneous decay increases slowly with a logarithmic dependence. This is the major reason that the obtained fidelity $\mathcal F$ has a weak dependence on the atom number $N$ instead of strong exponential dependence. It helps to extend the creation of GHZ states into the regime of thousands of atoms.

In Table.~\ref{stable2}, we find the optimized detuning $\Delta$ has a linear dependence on the cavity cooperativity $\eta$ for a given $N=100$. By fitting $\Delta/\Gamma$ versus $\eta$, we obtain the relation $\Delta/\Gamma=-0.185\eta$. This can be explained by the following arguments. When $\kappa$ and $\Gamma$ keep unchanged, $g^2$ is proportional to $\eta$. Since each atom in $\ket{\uparrow}$ introduces a frequency shift $\omega_s=g^2/\Delta=\eta\kappa\Gamma/(4\Delta)$, if we increase $\Delta$ but keep $\eta/\Delta$ as a constant, the frequency shift $\omega_s$ is kept the same for cavities with different cooperativities. The transmission property of the cavity is mainly determined by $\omega_s/\kappa$.
Therefore, the real part of the Hamiltonian becomes independent of the cavity cooperativity and the optimized trajectory of wavefunction evolution is almost the same for different $\eta$ when $\eta/\Delta$ is unchanged. Now the change of $\eta$ only affects the imaginary part and the fidelity increases with higher $\eta$ as higher $\eta$ results in less spontaneous decay.

\begin{table}[htp]
\caption{In Fig.~3\textbf{a} of the main text, we set $\eta=200$, $\kappa=2\pi\times 0.1$~MHz, $\Gamma=2\pi\times6$~MHz, and $\Delta=-36\Gamma$. We choose smaller $\Omega$ for the ensemble of larger atom number $N$ to keep $\sqrt{N}\Omega$ on the same order so that the matching light shift is reasonably small compared with detuning $\Delta$. Then we optimize other parameters to obtain the GHZ states with the highest fidelity.}
\begin{center}
\begin{tabular}{|p{1cm}|p{2.5cm}|p{1cm}|p{1cm}|p{1cm}|p{1cm}|p{1cm}|p{1cm}|p{2.5cm}|p{2.5cm}|p{1cm}|}
\hline
N		&$\Omega/(2\pi)$ (MHz)	&	$t_1$ ($\mu$s) 	&	$t_2$ ($\mu$s)	&	$t_3$ ($\mu$s)	&	$t_4$ ($\mu$s)	&	$\phi_3$	&$\phi_4$	&	$\delta_2$/(2$\pi$) (MHz)	&	$\delta_4$/(2$\pi$) (MHz)	&$\mathcal F$	 \\
\hline
100	&0.2	&	2.074	&0.285	&	0.191	&2.084	&0.503	&0.257	&-4.00	&-18.4	&0.924\\
\hline
200	&0.2 &	2.200	&0.200	&	0.135	&2.206	&0.503	&0.277	&-5.66	&-28.85	&0.919\\
\hline
300	&0.2	&	2.256	&0.163	&	0.110	&2.259	&0.503	&0.291	&-6.93	&-38.11	&0.916\\
\hline
400	&0.2	&	2.289	&0.141	&	0.095	&2.291	&0.503	&0.304	&-8.00	&-48.00	&0.913\\
\hline
500	&0.2	&	2.311	&0.126	&	0.085	&2.312	&0.503	&0.315	&-8.94	&-57.24	&0.911\\
\hline

600	&0.1	&	4.656	&0.231	&	0.156	&4.658	&0.503	&0.322	&-4.90	&-32.82	&0.909\\
\hline
700	&0.1	&	4.681	&0.214	&	0.144	&4.684	&0.503	&0.326	&-5.29	&-36.51	&0.907\\
\hline
800	&0.1	&	4.702	&0.200	&	0.135	&4.703	&0.503	&0.331	&-5.66	&-40.16	&0.905\\
\hline
900	&0.1	&	4.719	&0.188	&	0.127	&4.721	&0.503	&0.338	&-6.00	&-43.20	&0.903\\
\hline
1000	&0.1	&	4.733	&0.179	&	0.121	&4.735	&0.503	&0.342	&-6.32	&-46.80	&0.902\\
\hline
1100	&0.1	&	4.746	&0.170	&	0.115	&4.748	&0.503	&0.344	&-6.63	&-49.75	&0.900\\
\hline
1200	&0.1	&	4.757	&0.163	&	0.110	&4.758	&0.503	&0.346	&-6.93	&-52.65	&0.899\\
\hline
1300&0.1	&	4.766	&0.157	&	0.106	&4.767	&0.503	&0.349	&-7.21	&-55.53	&0.898\\
\hline
1400	&0.1	&	4.775	&0.151	&	0.102	&4.775	&0.503	&0.353	&-7.48	&-58.37	&0.896\\
\hline

1500	&0.05	&9.565	&0.292	&	0.197	&9.568	&0.503	&0.355	&-3.87	&-30.60	&0.895\\
\hline
1600	&0.05	&9.579	&0.283	&	0.191	&9.580	&0.503	&0.355	&-4.00	&-32.00	&0.894\\
\hline
1700	&0.05	&9.591	&0.274	&	0.185	&9.592	&0.503	&0.358	&-4.12	&-33.40	&0.893\\
\hline
1800	&0.05	&9.603	&0.266	&	0.180	&9.603	&0.503	&0.362	&-4.24	&-34.79	&0.892\\
\hline
1900	&0.05	&9.613	&0.259	&	0.175	&9.613	&0.503	&0.364	&-4.36	&-35.74	&0.891\\
\hline
2000&0.05	&9.623	&0.253	&	0.171	&9.626	&0.503	&0.364	&-4.47	&-37.12	&0.890\\
\hline
\end{tabular}
\end{center}
\label{stable1}
\end{table}%

\begin{table}[htp]
\caption{In Fig.~3\textbf{b} of the main text, we set $N=100$, $\kappa=2\pi\times 0.1$~MHz, $\Gamma=2\pi\times6$~MHz, $\Omega=2\pi\times 0.2$~MHz, $\delta_2=-2\pi\times 4$~MHz, and $\delta_4=-2\pi\times 18.4$~MHz unchanged for all the cases. We tune other parameters to optimize the performance of the final obtained GHZ states. We find the optimized rotated angles are the same for different $\eta$, but the optimized detuning $\Delta$ of the incident light is changing versus $\eta$ with a linear dependence.}
\begin{center}
\begin{tabular}{|c|c|c|}
\hline
$\eta$	&$\Delta/\Gamma$	&$\mathcal F$	 \\
\hline
100	&	-19	&0.867\\
\hline
200	&	-36	&0.924\\
\hline
300	&	-55	&0.944\\
\hline
400	&	-74	&	0.954\\
\hline
500	&	-92	&	0.956\\
\hline
600	&	-111	&	0.964\\
\hline
700	&	-130	&	0.966\\
\hline
800	&	-148	&	0.969\\
\hline
900	&	-167	&	0.970\\
\hline
1000	&	-186	&	0.972\\
\hline
\end{tabular}
\end{center}
\label{stable2}
\end{table}%

\section{The Fidelity of the obtained GHZ state versus $N$ and $\eta$}
In this section, we summarize an empirical formula for the fidelity $\mathcal F$ of the obtained GHZ state with respect to $N$ and $\eta$. In the previous section, we find the dissipation loss proportional to $\ln N$. We also find if we rescale the detuning $\Delta$ to keep $\eta/\Delta$ unchanged when $\eta$ increases,
the dynamic of the wavefunction evolution remains mostly the same except for the reduced dissipation loss which is inversely proportional to $\eta$.

\begin{figure}[htbp]
\begin{center}
\includegraphics[width=5.5in]{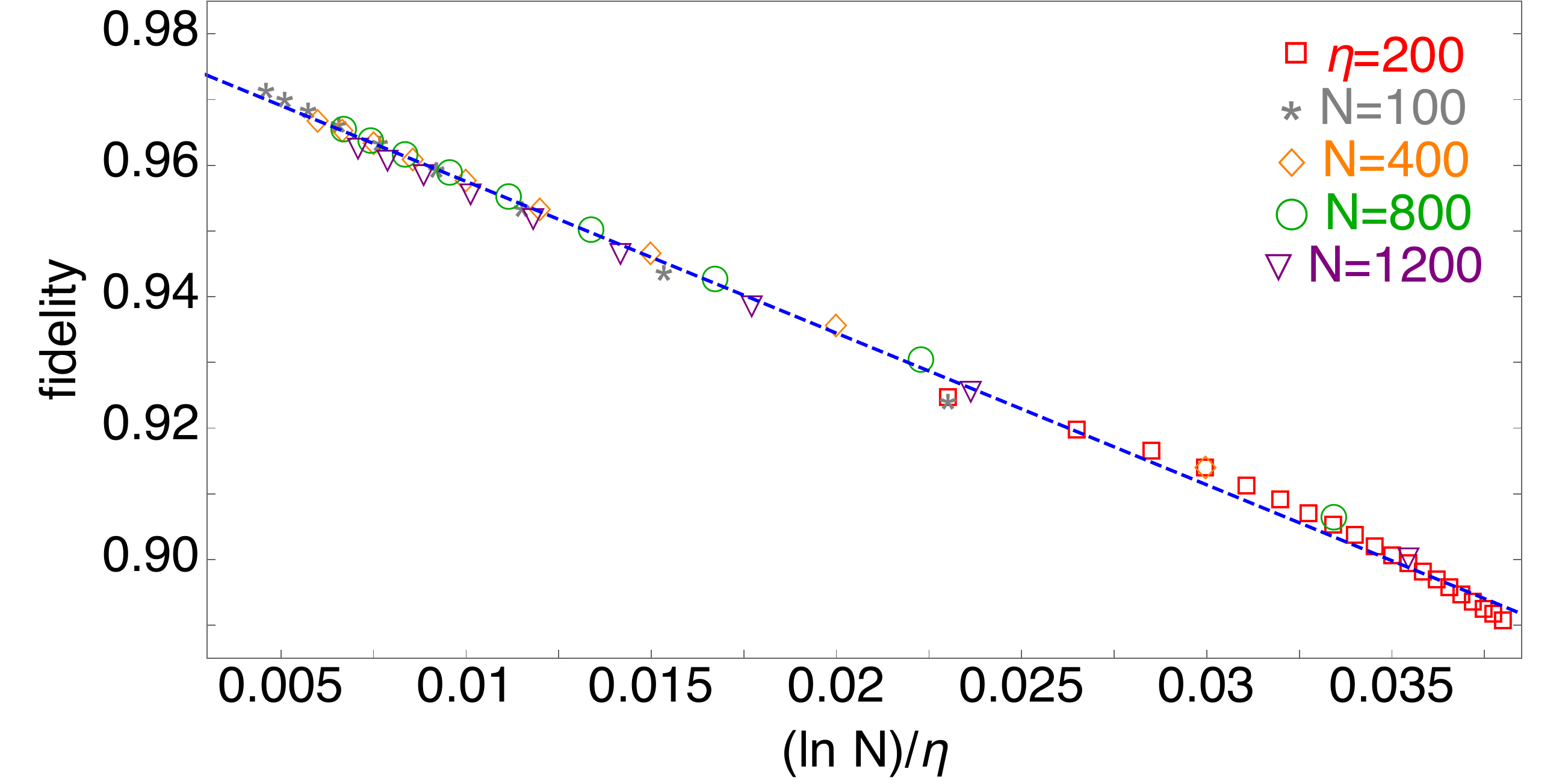}
\caption{Here we summarize five sets of different numerical results with total 56 points. The vertical axis is the obtained-GHZ-state fidelity $\mathcal F$ and the horizontal axis is $\eta^{-1}\ln N$. The red squares correspond to $\eta=200$ with different $N$ from 100 to 2000. The other four sets correspond to different $\eta$ but with a fixed $N$. The gray stars correspond to $N=100$, the orange diamonds correspond to $N=400$, the green circles correspond to $N=800$, and the purple triangles correspond to $N=1200$. $\eta$ varies from 200 to 1000. The blue dashed line is a linear fit $\mathcal F=0.981-2.31(\ln N)/\eta$.}
\label{sfig3}
\end{center}
\end{figure}

Therefore, by applying the same arguments we obtain the relation that $\mathcal F$ is linear depending on $\eta^{-1}\ln N$. In Fig.~\ref{sfig3}, we plot different sets of numerical results by varying $N$ and $\eta$. We find the data in a good agreement with the linear fit $\mathcal F=0.981-2.31(\ln N)/\eta$. The intersection is 0.981 lower than 1, and it is due to the imperfection of the cat-state creation in Step 2 which cannot be improved with a better cavity. The logarithmic dependence on $N$ extends our scheme to the regime with larger atom number. The difficulty of creating a giant GHZ state is then limited by the experimental noises such as the precision of Rabi rotations and cavity frequency stability. In the next section, we discuss that it is possible to create a 2000-atom GHZ state with technically achievable experimental noises.

\section{Robustness against common experimental noises}

In this section, we consider different kinds of experimental noises which may reduce the fidelity of the obtained GHZ state. 
We use rubidium 87 as the candidate atom, and two hyperfine state $\ket{F=2,m=0}$ and $\ket{F=1,m=0}$ as the candidate levels, for they have long coherence time and are not sensitive to the fluctuation of magnetic field. The transition between these two levels has been successfully used for the atomic clock, in which the coherence time can be longer than a second. Thus, it is a reasonable assumption that the coherence time of a 2000-atom GHZ state is at least on the order of milliseconds.
According to the previous section, all the steps can be achieved within microseconds. Therefore, the scheme is robust against technical noises that decrease the coherence time between $\ket{\uparrow}$ and $\ket{\downarrow}$.

Here, we discuss effects of common noises existing in the laboratory, including Rabi rotations, inhomogeneous coupling, and the frequency fluctuation. The conclusion is that \textit{entanglement amplification} is robust against these noises and secures the realization of GHZ states with high fidelity claimed in the main text.

\subsection{The precision of Rabi rotations}

In this subsection, we consider the errors from Rabi rotations. We use RF drivings to control the Rabi rotation angle under the
precise control of the RF power and pulse duration that are available in most of the cold atom labs.

The timing control error can be suppressed below 1~ns by simple synchronization with a rubidium frequency standard, so we can use 1~ns as the time fluctuation for the following estimations. Besides, we assume the shot-to-shot intensity fluctuation is 0.4\% which leads to an amplitude fluctuation at 0.2\%. Assuming both kinds of noise satisfy the Gaussian distribution, and for each step, we pick up a random $\delta t$ and a random $\delta\Omega/\Omega$ both centered at 0 and with a standard deviation at 1~ns and 0.2\% respectively, and then use these two numbers to correct corresponding time $t_i$ and Rabi frequency $\Omega$.

After correcting all the steps for creating the GHZ state, 
we obtain a final state $\ket{\psi_f}$. For each particular atom number,we repeat these random corrections for more than 200 times and then calculate the average density matrix $\rho=\overline{|\psi_{f}\rangle\langle \psi_{f}|}$ which represents the actual obtained states under the influence of these random noises. We find the fidelities decrease to 0.923 ($N=100$), 0.904 ($N=500$), 0.890 ($N=1000$), and 0.870 ($N=2000$) while the original values are 0.924 ($N=100$), 0.911 ($N=500$), 0.902 ($N=1000$), and 0.890 ($N=2000$).

\subsection{Inhomogeneous coupling and photon shot noise}

The inhomogeneous coupling and photo shot noise affect the actual AC Stark shift. Both of them  
introduce a fluctuation of the light shift $\delta$ in $H'_{\text{exp}}$. If we use commensurate wavelength lasers such as 1560~nm and 780~nm for trapping and probing rubidium atoms, we can avoid the inhomogeneous coupling by introducing standing waves. The remained inhomogeneity is then due to thermal fluctuations. The broadened linewidth of atoms with a temperature at 10~$\mu$K trapped by a standing wave with a trap depth at 20~MHz only causes a random reduction of the coupling strength by 0.5\%, \textit{i.e.} 1\% reduction for AC Stark shift. Here we use $(1-\epsilon)\delta$ to represent the random reduction of AC Stark shift. Let's assume the mean intra-cavity photon number is $\bar{n} $ for the following analysis. 

We apply the random-number method to calculate the state evolution. Each time we pick up a random $n$ and a random $\epsilon$, where $n$ is a positive integer and Poisson-distributed with a center at $\bar{n}$ and $\epsilon$ satisfies a half Gaussian distribution with a center at 0 and a squared mean $\langle\epsilon^2\rangle$ at $(1\%)^2$. Then we correct the light shift $\delta$ by $n(1-\epsilon)\delta /\bar{n}$. For Step 2 and 5, we use different random numbers to perform the corrections. After 200 repetitions, we obtain the average density matrix $\rho$ and calculate the fidelity $\mathcal F$. We find the fidelities decrease to 0.904 ($N=100$), 0.903 ($N=500$), 0.891 ($N=1000$), and 0.881 ($N=2000$) while the original values are 0.924 ($N=100$), 0.911 ($N=500$), 0.902 ($N=1000$), and 0.890 ($N=2000$).

\subsection{The instability of frequency}

The frequency fluctuations can be considered in two parts. One is common frequency fluctuations where the frequency difference between the cavity and the laser keeps unchanged, the other is the fluctuation of their frequency difference. As for the former, the absolute frequency of a laser-cavity system can be easily stabled below 1~MHz. Compared with the detuning $\Delta=-36\Gamma=-2\pi\times 216$~MHz, 1~MHz collective frequency fluctuation only introduces $0.5\%$ instability to AC Stark shift. 

The major issue here is the latter one where the frequency difference between the cavity and the laser fluctuates. In most cavity experiments, one can use the method of transfer cavities or frequency doubling to suppress such relative frequency fluctuations to within 0.2$\kappa$. Therefore, we assume the frequency jittering to be $0.2\kappa$.

By applying the same arguments and calculations in last subsection, we find the fidelities decrease to 0.920 ($N=100$), 0.881 ($N=500$), 0.856 ($N=1000$), and 0.831 ($N=2000$) while the original values are 0.924 ($N=100$), 0.911 ($N=500$), 0.902 ($N=1000$), and 0.890 ($N=2000$).

\subsection{Overall robustness against noise}
By integrating all the considerations above, we find the fidelities decrease to 0.903 ($N=100$), 0.867 ($N=500$), 0.845 ($N=1000$), and 0.817 ($N=2000$), while the original values are 0.924 ($N=100$), 0.911 ($N=500$), 0.902 ($N=1000$), and 0.890 ($N=2000$). These calculations and arguments confirm the robustness of our scheme and support that the scheme is experimentally achievable. The main reason for the robustness is that our scheme is relying on the off-resonant suppression which tolerates a wide range of parameters.

\begin{figure}[htbp]
\begin{center}
\includegraphics[width=5.5in]{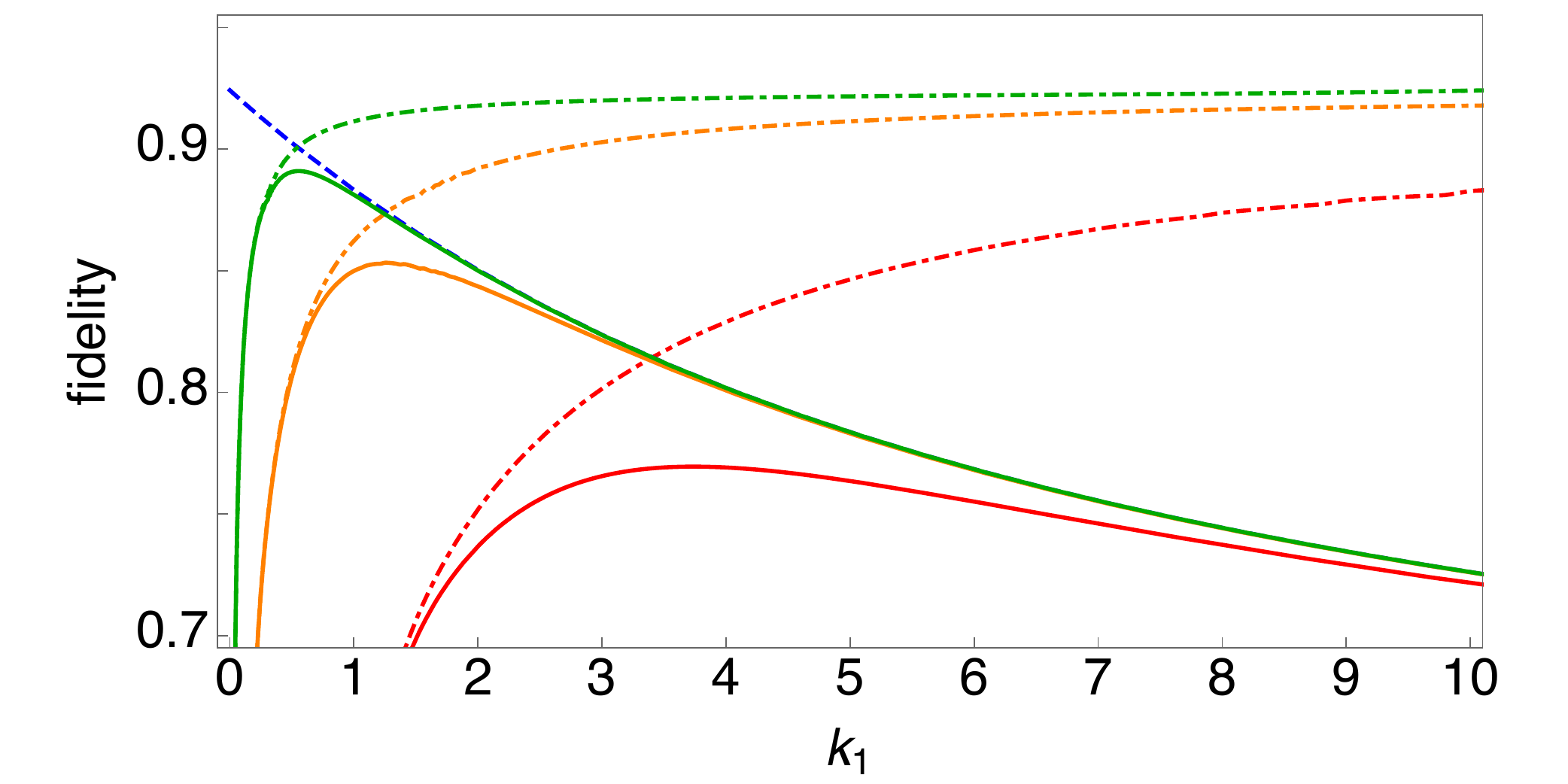}
\caption{The fidelity versus $\bar k_1$ under $N=100$ and $\eta=200$. The blue dashed line corresponds to the decoherence from the information leakage. The red, orange, and green dot-dashed lines correspond to the influence by photon shot noise under different cavities. The red one describes a symmetric cavity, the orange one describes an asymmetric cavity with 5 ppm and 59 ppm transmission respectively, and the green one describes an asymmetric cavity with 1 ppm and 63 ppm transmissions. The red, orange, and green solid lines correspond to the overall fidelity integrating both the information leakage and the photon shot noise.}
\label{sfig4}
\end{center}
\end{figure}

\section{Information leakage}
The incident light is transmitted through the cavity and the transmission ratio depends on the atomic state. So the transmitted photon carries the information of the atomic state and serves as a measurement. If we go through the GHZ state creation procedure described in the main text, with 100 atoms in a symmetric cavity, there are 28.8(or 3.5) photons transmitted on average if the atoms are at the state $\ket{m=-N/2+1}$(or $\ket{m=-N/2+2}$). 
These photons would collapse the atomic wave function mostly into $\ket{m=-N/2+1}$ with small population into $\ket{m=-N/2}$ and $\ket{m=-N/2+2}$. Such a process decreases the fidelity of the obtained GHZ states.

We estimate the reduction of fidelity for a symmetric cavity as follows. If the atomic state is at $\ket{-N/2+n}$, the mean transmitted photon number $\bar k_n$ is proportional to the transmission spectra $|T(\xi,n)|^2$. Here we use the mean transmitted photon number $\bar k_1$ for $\ket{-N/2+1}$ to characterize all the other states, then
\begin{equation}
\bar k_n=\bar k_1 |T(\xi,n)|^2/|T(\xi,1)|^2. 
\end{equation}
The probability 
to detect $k$ transmitted photons for the state $\ket{-N/2+n}$ is labeled as
$p(k,n)=\bar k_n^k\exp(-\bar k_n)/k!$.
Thus, we define a positive-operator-valued measure
$\{E^\dagger (k) E(k)\}$ where $E(k)=\text{diag}\{\sqrt{p(k,N-i+1)}\}$ with the matrix index $i$ from 1 to $N+1$. 
The information leakage happens in Step 2 where the transmitted photons disturb the phase information of the atomic state. The actual state $\rho_2$ in Step 2 is described by a density matrix
\begin{equation}
\rho_2=\sum_{k=0}^\infty E(k)|\text{Step 2}\rangle\langle \text{Step 2}|E^\dagger(k)
\end{equation} 
On the one hand, when using the updated $\rho_2$ to generate the final state and calculate the fidelity, we find the fidelity decreases as $\bar k_1$ increases. Thus, it is preferable to choose a smaller $\bar k_1$ to reduce the harm of information leakage.
On the other hand, a smaller $\bar k_1$ corresponds to a larger relative photon shot noise and we can characterize the shot noise by the same method in the previous section.
In Fig.~\ref{sfig4}, we plot the fidelity $\mathcal F$ versus $\bar k_1$ for the case of $N=100$ with $\eta=200$. The red solid line is the overall fidelity integrating both the effects of shot noise and information leakage. The optimized fidelity is 0.769, while the original value is 0.924.

The reduction of fidelity due to information leakage in a symmetric cavity is significant. However, this problem can be solved by using a single-side cavity. If one of two mirrors is a perfect-reflection mirror, all the photons will be reflected and no information leakage will occur.
However, it is super hard to obtain a perfect-reflection mirror.
In Ref. \cite{PhysRevA.85.013803}, an incident light is coherently-divided by a 50:50 beam splitter, then two parts of the light are sent into both sides of a symmetric cavity instantaneously. Transmission and reflection exist at the same time in both sides of the cavity, which makes it impossible to distinguish whether a photon is reflected or transmitted.
This erases the information carried by the transmitted photons, thus avoids destructing the GHZ state.

Another alternate method is by using an asymmetric cavity, where one mirror has a higher transmission amplitude $q_1$ and the other has a lower transmission amplitude $q_2$. The photons are sent from $q_1$ side. Due to the asymmetry, assuming there are $n$ atoms in $\ket{\uparrow}$ coupled with the cavity field and following the same derivations in Ref. \cite{TANJISUZUKI2011201}, we construct the transmission amplitude by
\begin{equation}
\tilde{T}(\xi,\eta)={2q_1 q_2\over q^2_1+q^2_2}{1\over 1+ {n\eta\over 1+4(\Delta+\xi)^2/\Gamma^2}-2i\left[{\xi\over\kappa}-n\eta {(\Delta+\xi)/\Gamma \over 1+4(\Delta+\xi )^2/\Gamma^2}\right]}={2q_1 q_2\over q^2_1+q^2_2}T(\xi,\eta).
\end{equation}
The intra-cavity field $E_c$ is proportional to 
\begin{equation}
E_c\propto{2 q_1\over q^2_1+q^2_2}{1\over 1+ {n\eta\over 1+4(\Delta+\xi)^2/\Gamma^2}-2i\left[{\xi\over\kappa}-n\eta {(\Delta+\xi)/\Gamma \over 1+4(\Delta+\xi )^2/\Gamma^2}\right]}.
\end{equation}
With the same cooperativity $\eta$ and the same intra-cavity power, the transmission amplitude of an asymmetric cavity is reduced by a factor of $\sqrt{2 q^2_2/(q^2_1+q^2_2)}$, and the photon transmission rate is reduced by a factor of ${2 q^2_2/(q^2_1+q^2_2)}$.

Now let's estimate the reduction of fidelity resulted from information leakage with experimentally available parameter. For a cavity with $\eta=200$
and finesse $=100000$, the waist of the cavity mode is around 7.7~$\mu$m and the total transmission ratio of both mirrors is 64 ppm. For a symmetric cavity, each mirror has 32 ppm transmission, and the transmission ratio on $\ket{m=-N/2+1}$ is 1.
For an asymmetric cavity, one mirror could have 5 ppm transmission while the other has 59 ppm (orange lines in Fig.~\ref{sfig4}), and the on-resonance transmission ratio becomes 0.288. Thus, under the same intra-cavity intensity, the photon transmission ratio is reduced by a factor of 0.156 when using the asymmetric cavity mentioned above. Therefore, the mean transmitted photon number for the atomic state $\ket{m=-N/2+1}$ (or $\ket{m=-N/2+2}$) becomes 4.49 (or 0.55), and the final fidelity of the GHZ state becomes 0.853.
This could be further improved by a more aggressive asymmetric design, such that a cavity with the parameters of 1 and 63 ppm (or 3 and 61 ppm) results in a fidelity of 0.891 (or 0.867)(green lines in Fig.~\ref{sfig4}).

When there is loss in the cavity system such as the defects on the mirrors or the mode clipping, we can count all the loss into the smaller transmission ratio $q_2$, which gives the worst scenario in the asymmetric-cavity case. To our knowledge, the loss is mainly caused by the surface roughness and coating quality of the cavity mirror, and could be controlled below 1 ppm by the state-of-art fabrication. Thus, the mirror transmission of 5ppm and 59 ppm is a reasonable estimation for asymmetric cavities.

\end{document}